\documentclass[aps,prb,twocolumn,superscriptaddress]{revtex4-1}
\usepackage{hyperref} 
\usepackage{graphicx}
\usepackage{amsmath}
\usepackage{amssymb}

\usepackage{xcolor}

\begin{document}

\title{Defect-induced magnetism and Yu-Shiba-Rusinov states in twisted bilayer
graphene}

\author{Alejandro Lopez-Bezanilla}
\email[]{alejandrolb@gmail.com}
\address{Theoretical Division, Los Alamos National Laboratory, Los Alamos, New Mexico 87545, USA}
\author{J. L. Lado}
\address{Institute for Theoretical Physics, ETH Zurich, 8093 Zurich, Switzerland}

\begin{abstract}
	Atomic defects have a significant impact in the low-energy properties of graphene systems.
	By means of first-principles calculations and tight-binding models we provide evidence that chemical impurities modify both the normal and the
        superconducting states of twisted bilayer graphene.
A single hydrogen
	atom attached to the bilayer surface yields a triple-point crossing, whereas self-doping and
	three-fold symmetry-breaking
	are created by a vacant site. 
Both types of defects lead to time-reversal
	symmetry-breaking and the creation of local magnetic moments.
	Hydrogen-induced magnetism
	is found to exist also at the doping levels
	where superconductivity appears in magic angle graphene superlattices.
As a result, the coexistence of superconducting order and defect-induced magnetism yields
	in-gap Yu-Shiba-Rusinov excitations in magic angle twisted bilayer graphene. 
\end{abstract}
 
\date{\today}

\maketitle

\section{\label{sec:intro} Introduction}

The tunability of graphene\cite{RevModPhys.81.109} 
and two-dimensional materials has provided
an outstanding solid-state platform to explore emergent physical phenomena.
Yet, on top its standalone interest,
graphene provides a powerful building block
to create superstructures due to
the weak van der Waals forces between layers.\cite{Geim2013}
Those very same
weak van der Waals forces
allow to deposit graphene layers on top of each other with relative
angles, in stark contrast with conventional
bulk crystals.
Twisted bilayer graphene (tBLG)\cite{dosSantos} is one the simplest
structures that can be built in that fashion,
and has opened the door towards
realizing states of matter inaccessible in monolayer
graphene.
\cite{Rickhaus2018,Cao2018,Cao2018correlated,2019arXiv190103520S,2019arXiv190103710C,2019arXiv190410153J}
This additional flexibility stems
from the emergent moir{\'e}
pattern
that arises due to the twist between the layers, giving rise
to an emergent electronic structure that
can be controlled by the twist angle.\cite{PhysRevB.82.121407,Bistritzer2011,PhysRevB.88.121408,PhysRevLett.121.146801}

A paradigmatic
example of new physics
associated with the emergent moir{\'e}
band structure
is the appearance of superconductivity
in magic angle graphene ($\alpha \approx 1^\circ$)
superlattices,\cite{Cao2018} 
that exploit a chemical-free field effect doping
of graphene without altering its structural integrity.
Moreover,
experimental observations reported strong correlated
behavior,\cite{Cao2018correlated} 
anomalous Hall
effect,\cite{2019arXiv190103520S} strange metal
behavior\cite{2019arXiv190103710C}, and rotational
symmetry-breaking\cite{2019arXiv190410153J} close
to that regime.
Extensive theoretical efforts are being directed
towards deriving a faithful low-energy model,\cite{PhysRevX.8.031088,PhysRevX.8.031087,PhysRevX.8.031089,Guinea2018}
studying the possible
electronic instabilities,\cite{PhysRevX.8.041041,PhysRevLett.121.087001}
and providing means of tailoring
the electronic properties
of such state.\cite{Chittari2018,Yankowitz2019,PhysRevB.98.085144}
However, the influence of atomic defects in this
structures has not been addressed in detail, and their potential
impact in the low-energy properties
has remained rather unexplored.\cite{Brihuega2017,2019arXiv190205862R}

\begin{figure}[!t]
 \centering
         \includegraphics[width=0.48\textwidth]{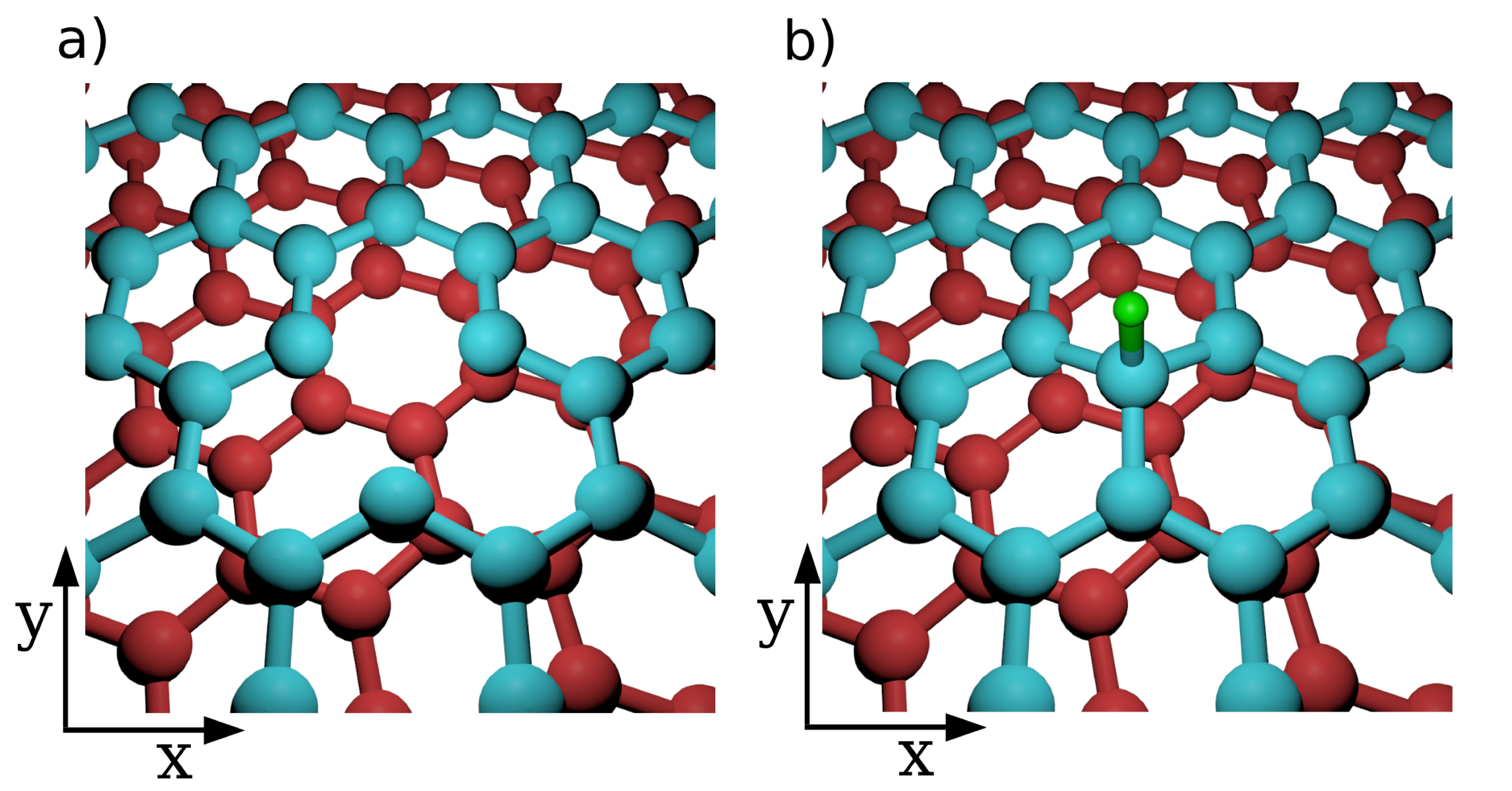}
 \caption{ Sketch of the two defects
	considered in tBLG,
	a carbon vacancy a) and
	a chemisorbed hydrogen atom b).
	Blue denotes the upper layer, red the lower layer
	and green the hydrogen atom.
	Both defect create impurity bands close the
	charge neutrality point
	that give rise to
	local magnetic moments,
	substantially impacting
	the low-energy properties of twisted bilayers.}
 \label{fig:sketch}
\end{figure}

In this paper, we study the impact of atomic defects
in tBLG with angles ranging from
large angles $\alpha\approx 5^\circ$
to the magic angle
$\alpha \approx 1^\circ$.
First-principles methods and
effective real-space
tight-binding models are used to describe two paradigmatic
defects in graphene systems, namely
carbon vacancies\cite{PhysRevLett.104.096804,Yazyev2010,PhysRevB.77.195428,PhysRevB.77.115109,PhysRevB.75.125408}
and chemisorbed H
atoms\cite{Brihuega2017,GonzalezHerrero2016,PhysRevB.96.024403,LopezBezanilla2009,2019arXiv190205862R}
as shown in Fig. \ref{fig:sketch}.
In particular, the chemical modification introduced by mono-hydrogenation
creates flat bands at the Fermi energy, whose 
unbalanced electronic occupation
yields local magnetic moments.
In the superconducting
state of magic-angle tBLG,
such magnetic moments are shown to
in-gap Yu-Shiba-Rusinov 
excitations, dramatically impacting the
spectral properties of the superconducting state.

The manuscript is organized as follows.
In Sec. \ref{sec:dft} 
the emergence of magnetism as a result of both C vacancies and mono-hydrogenation is analyzed with first-principles calculations.
Sec. \ref{sec:lowenergy}
shows that chemisorbed H
can be treated with
a low-energy model that reproduces accurately 
the first-principles results.
In Sec. \ref{sec:shiba} we show that the interplay between
impurity induced magnetism
and superconductivity gives
rise to in-gap Yu-Shiba-Rusinov states.
Finally, Sec. \ref{sec:conclusions} summarizes our conclusions.

\section{\label{sec:RandD}Defect-induced states from first principles}
\label{sec:dft}
\subsection{General considerations}

\begin{figure}[!t]
 \centering
	 \includegraphics[width=0.5\textwidth]{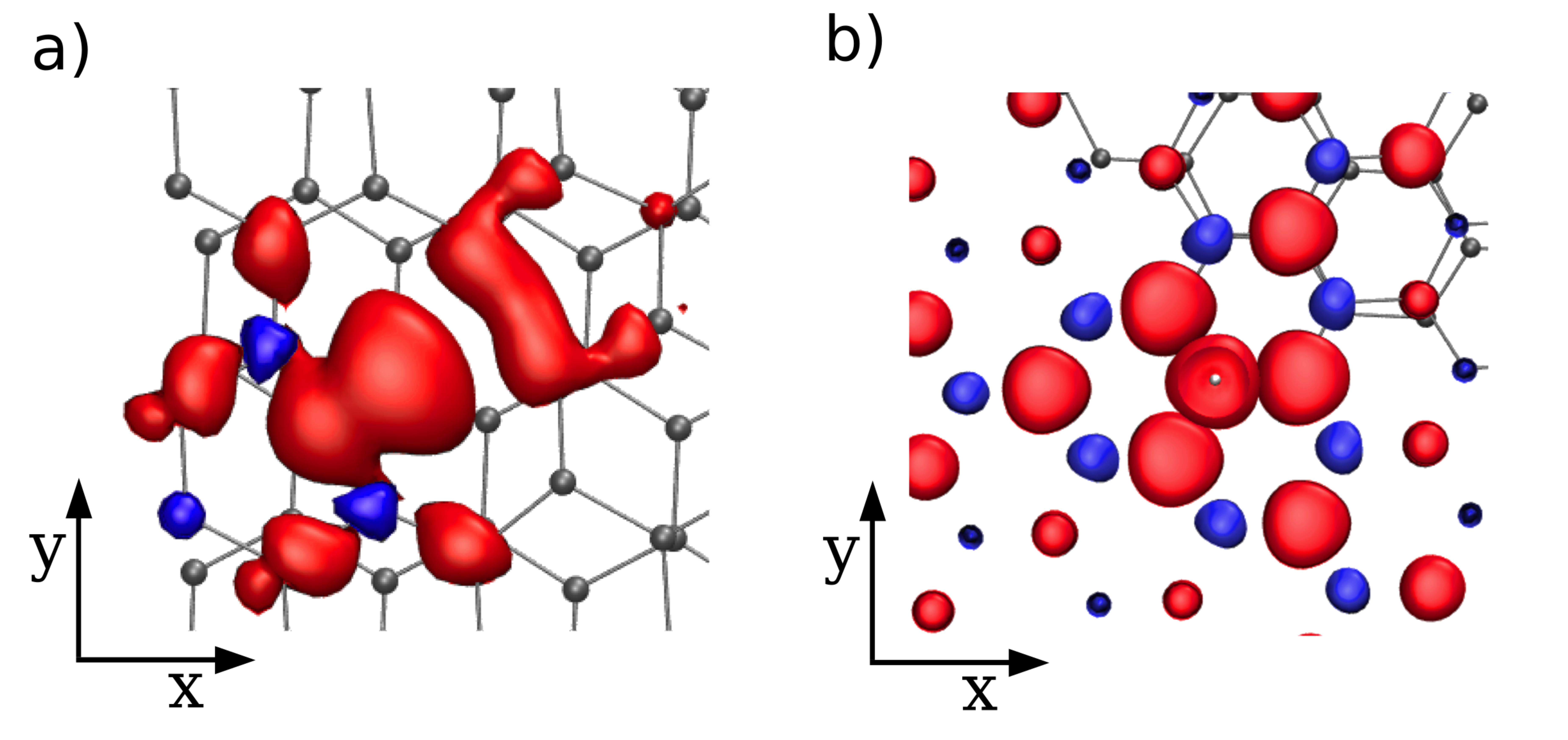}
 \caption{Real-space redistribution
	of the charge density of two
	different defects
	in twisted
	bilayer graphene at $\alpha = 5.09^\circ$.
	Top view of the charge distribution of a single vacancy, a) and a
	single H atom, b).
	A single-H atom creates a triangular distribution around the defect,
	whilst the vacant site enhances the net charge localization at the
	$\sigma$ dangling bonds. 
	Isosurfaces correspond to charge densities of $10^{-3}$ e$^-$\AA$^{- 3}$( a), b)) and $10^{-2}$
	e$^-$\AA$^{-3}$ (c))}  
 \label{figRealSpace}
\end{figure}

We first study the effect of structural
and chemical modification of tBLG 
with first-principle methods. 
Moir{\'e} lattices are created by twisting a graphene layer over the other by a
commensurate angle, so that a unit cell with coincidence site lattice points is
created\cite{Campanera,dosSantos,Shuyang}. 
In the chosen tBLG geometry, one layer of AA-bilayer graphene is
rotated an angle $\theta= 5.09^\circ$ with respect to the other layer. Internal
coordinates of pristine and modified tBLGs separated initially by the van der
Waals distance of 3.3\AA\ were fully relaxed. 
The momentum mismatch between the layers
yields two linearly dispersive bands,
which are degenerate
in the $\Gamma - K - M$ high-symmetry
path of the 
BZ.\cite{PhysRevB.82.121407,dosSantos,Bistritzer2011,Latil2006,Latil2007}

First-principles calculations provide
a powerful insight on effect of
structural relaxation on the electronic structure of defective
systems, together with
the properties of the remaining unpaired electron. 
Chemisorption of a single H atom \cite{Brihuega2017,GonzalezHerrero2016,PhysRevB.96.024403,LopezBezanilla2009,2019arXiv190205862R}
is a simple way of creating an imbalance
between the number of A and B sites in
graphene,
with bare modification of the bipartite
nature of graphene
lattice\cite{PhysRevLett.113.246601,PhysRevLett.121.136801,PhysRevLett.107.016602,PhysRevLett.110.246602,PhysRevLett.114.246801,PhysRevLett.112.116602,PhysRevB.98.155436,PhysRevB.83.193411,PhysRevB.88.085441,PhysRevB.81.075423,PhysRevLett.113.246601,PhysRevB.77.134114}.
The imbalance between the number of sites
in the two sublattices has an important
consequence that follows from Lieb's theorem\cite{PhysRevLett.62.1201}:
a bipartite lattice with more
A sites than B sites will show a number of zero
modes proportional to such difference.
In a similar fashion, 
a vacant C site breaks the lattice symmetry by removing one
atom, but creating addtional distortions\cite{PhysRevLett.101.037203,PhysRevLett.117.166801,PhysRevLett.93.187202,PhysRevB.85.115405,PhysRevB.99.125125,PhysRevB.94.075114,PhysRevB.86.165438}.
In an oversimplified picture, both approaches are equivalent to removing a
$\pi$ electron. However, the sp$^3$ rehybridization of an atom or the creation
of
dangling $\sigma$-bond have completely different consequences on the
localization an dispersion of the electronic state. This feature
can be clearly seen in the charge redistribution of Fig. \ref{figRealSpace},
where it is shown that 
the vacancy triggers an atomic reconstruction
that breaks $C_3$ symmetry (Fig. \ref{figRealSpace}a),
whereas
the hydrogen ad-atom creates
a redistribution that
conserves the original three-fold rotation
(Fig. \ref{figRealSpace}b).

Due to the varying distance between C atoms across the parallel layers, the
effect of a H atom depends on its
location.\cite{PhysRevB.85.205402,PhysRevLett.121.136801,PhysRevB.82.235409,PhysRevB.83.165402,PhysRevB.77.115114}
Similarly, a vacant site interacts differently with the opposite graphene layer
depending on the stacking region where the C atom was removed
from.\cite{PhysRevB.85.245443,PhysRevMaterials.2.034004,PhysRevB.89.245429}
In the following, we address in detail the electronic
reconstruction associated to each one of the defects introduced.

\subsection{Carbon vacancy}
\begin{figure}[!t]
 \centering
	 \includegraphics[width=0.48\textwidth]{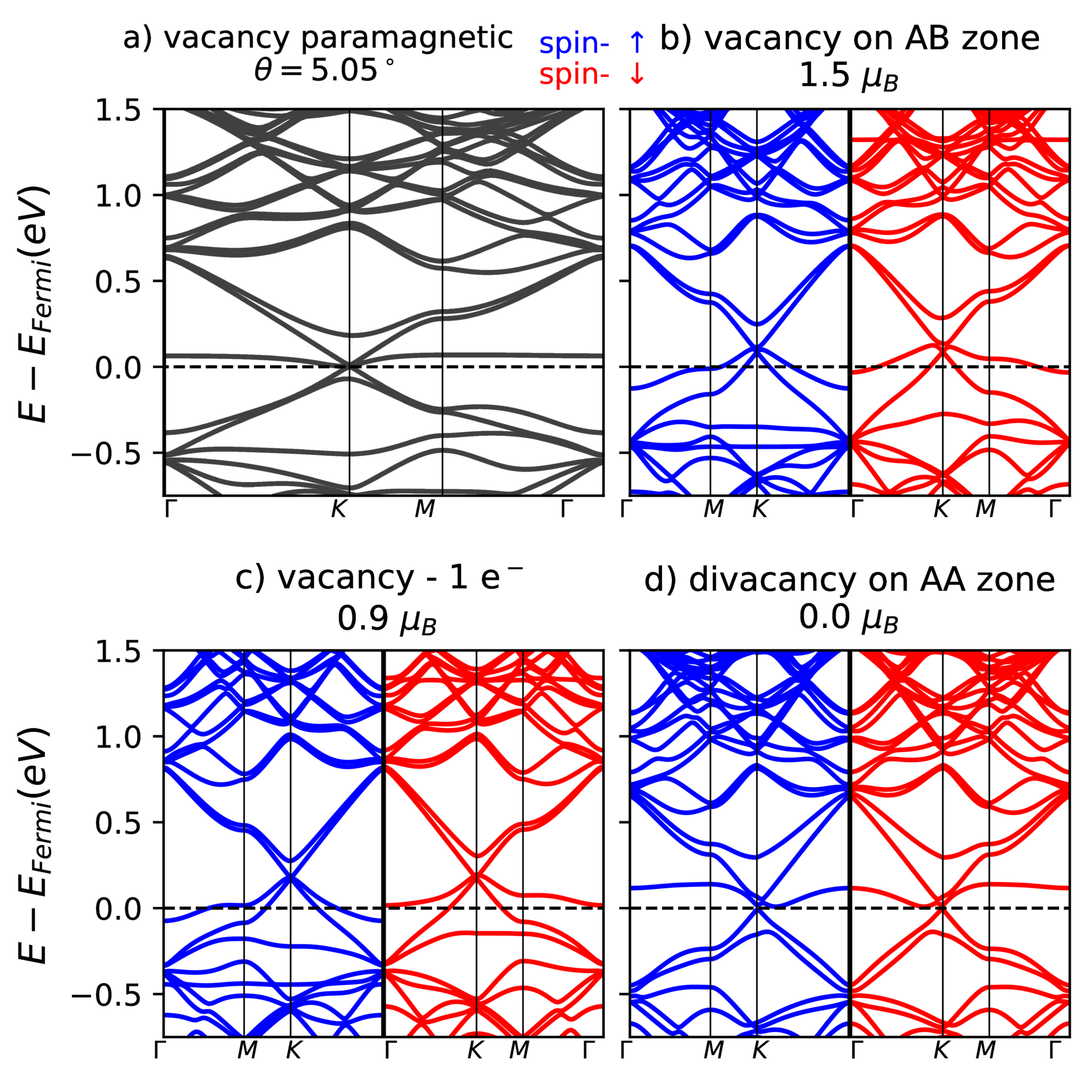}
 \caption{a) 
	First-principles
	paramagnetic electronic band structure of a bilayer graphene with a
	twist angle of $\alpha = 5.09^\circ$
	and a C atom vacancy. Spin-polarized
	calculations yield a spin splitting of the defect-induced band.
	In b), a vacancy in the AB stacking zone yields a
	magnetic moment of 1.5 $\mu_B$, which is lowered to 0.9 $\mu_B$ when
	removing one electron. A di-vacancy
	exhibits a paramagnetic configuration and an empty nearly flat-state at
	Fermi level. In all four cases the impurity state exhibits at $\Gamma$
	point an energy in between the to Dirac cones.  } 
 \label{figvac}
\end{figure}

Within the DFT formalism, 
a vacancy is more than a simple removal
of a lattice site, and reconstruction of the whole tBLG structure is
considered\cite{PhysRevLett.104.096804}. Atomic rearrangement is present in
real systems, and thus
removing a C atom out of one of the graphene sheet has a
non-trivial effect
on the distribution of electronic states in the low-energy band diagram. Fig.
\ref{figvac}a displays a paramagnetic calculation of a fully relaxed tBLG
structure with a vacant site. The electronic state at the Fermi level
corresponds to the defect state, which removes the degeneracy of the bands
that form the Dirac cones and creates an almost degenerate (4 meV gaped)
triple-crossing at the $K$ point. The vacancy state creates a major disruption
on one of the sets of linear bands, pushing them at higher and lower energies.

\begin{figure}[!t]
 \centering
	 \includegraphics[width=0.48\textwidth]{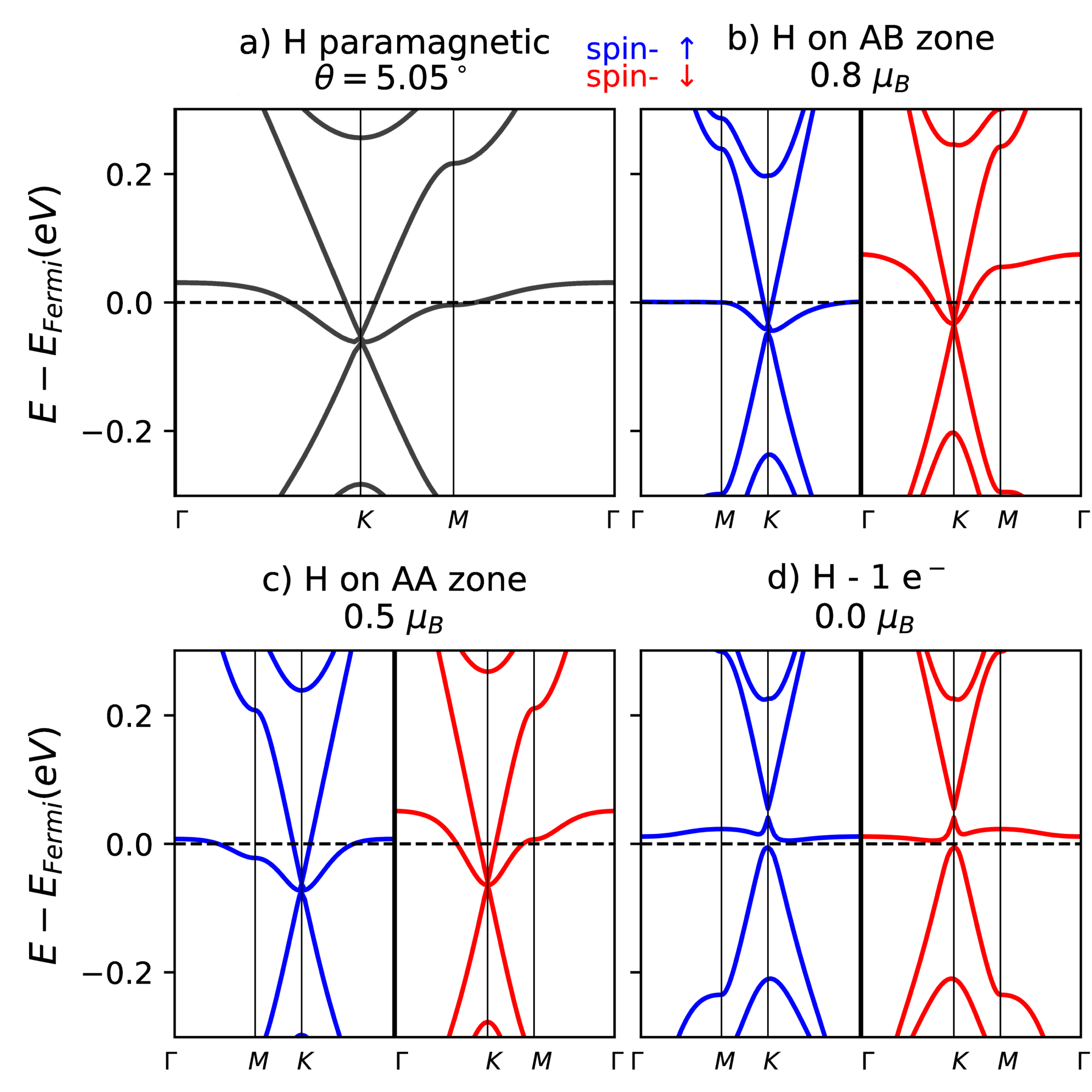}
 \caption{a) First-principles
	paramagnetic band structure of mono-hydrogenated
	$\alpha \approx 5 ^\circ$
	tBLG. 
	When an H atom is attached to a C atom in the AB b) and AA c)
	regions, a local magnetic moment arises in the system,
	lifting the original spin degeneracy. 
	Panels
	b) and c) show the unbalanced occupation of the
	nearly nearly flat band in the up/down channels,
	yielding a local magnetic moment
	mentioned above. In this large angle regime
	$\alpha \approx 5 ^\circ$,
	removing one electron out of
	the system leads to a non-magnetic and gaped configuration, with an
	empty electronic
	flat-state in the vicinity of the Fermi level.    }
 \label{figH}
\end{figure}

By introducing the spin degree of freedom,
a magnetic ground state
with an associated
spin splitting of the defect band
is found
(Fig.\ref{figvac}b).
A magnetic moment of 1.5 $\mu_B$ is created as a result of
the p-type doping character that unbalances the $\pi$-bands filling at the
Fermi level. 
A vacancy
in a tBLG polarizes the bilayer by distorting the dispersion of the $\pi$-bands
and introducing a p-doping on the opposite layer that unbalances the electronic
population of the valence bands. Furthermore, the defect creates two
spin-dependent dispersionless bands at very different energies, (-0.5 and 1.35 eV for spin-$\uparrow$ and spin-$\downarrow$ respectively) as show in Fig.\ref{figvac}b, and whose
spatial extension is reduced to the immediate surrounding of the vacancy. Real-space representation of the charge density
distribution in Fig.\ref{figRealSpace}a 
shows that
the unpaired electron is located in the $\sigma$-bonds of the C atoms at
the vacant site.
The highly localized nature of the defect, together with the
location of its electronic states in the band diagram is expected to yield
a greatly enhanced signal in a scanning tunnel microscope
experiment.\cite{Brihuega2017,PhysRevLett.104.096804} 

Removing one electron out of the system barely changes the shape of the
electronic bands, but empties completely the spin-$\uparrow$ band, leaving the
other partially filled (see Fig.\ref{figvac}c). A total magnetic moment of 0.9
$\mu_B$ is obtained, similarly to the  case of monolayer
graphene\cite{PhysRevLett.104.096804}. Removing an additional C atom
near to the
removed site creates a divacancy, that restores
the non-magnetic state of the structure
by effectively removing
the unpaired electrons of the single-vacancy. As shown in Fig.
\ref{figvac}d, a defect state similar to the paramagnetic single vacancy state
is located in the vicinity of the Fermi level and the flat bands at low and
high energies disappear.

\subsection{Chemisorbed hydrogen atom}

We now move on to consider the effect of a single H atom attached to one of the
graphene layers. Relaxation of the entire structure results on the
adsorbent C atom site pulled out of the graphene plane, adopting the pyramidal
geometry characteristic of the sp$^3$ hybridization. Moreover,
the bond lengths between the
anchoring carbon atom and its first neighboring atoms were elongated to 1.49
\AA, and the C-H bonding distance is of 1.14 \AA. 
Interestingly, the atomic reconstruction
preserves the local $C_3$ symmetry (Fig. \ref{figRealSpace}b),
in strike comparison with the
vacancy case.
The atomic positions in the
opposite layer are barely affected,
and the band structure of the system
shows a triple point crossing
at the K point (Fig. \ref{figH}a).
In a spin polarized calculation,
a spin splitting appears in the band structure,
breaking time reversal symmetry
as a consequence of the
net magnetic moment triggered by
the presence of the hydrogen zero mode
(Fig. \ref{figH}bc).
In particular, the exchange field
generates 
a shift of $\sim$0.25 meV
up and down of the impurity band.
Interestingly,
for any attachment position, the
linear dispersion typical of graphene is
preserved,
although the Dirac cones are
shifted in energy and a meV gap removes the Dirac point. 

For an H atom sitting in the AB stacking region, one of the electronic bands
associated with the defect is almost fully populated with one electron, whereas
a similar band with opposite spin quantum number is almost empty. The total
magnetic moment is of 0.8 $\mu_B$. A similar behavior is observed for an H atom
sitting on a AA stacked C atom, although the resulting magnetic moment
decreases to 0.5 $\mu_B$. A small n-type doping is induced in the layer
opposite to the one with the H atom. This last feature
can be inferred from the band diagrams of
Fig.\ref{figH}b and c, where the upper Dirac cone becomes slightly filled. This
electron donor character demonstrates that both layers are partially hybridized
through the defect state and a partial charge transfer occurs between layers.
Finally, we consider the effect of a doping of a single electron
per moir{\'e} supercell in this regime
$\alpha \approx 5^\circ$. In that situation,
we observe
that the magnetic moment
of the impurity state
is completely quenched (Fig.\ref{figH}d),
yielding a fully occupied impurity band.
As it will shown in Sec. \ref{sec:shiba},
the phenomenology at $\alpha \approx 1^ \circ$
is dramatically different
due to the appearance of the
magic angle flat bands.

The previous phenomenology applies to twisted bilayers
with large twisting angle $\alpha \approx 5^\circ$. In the following
we will focus in smaller angles tBLG, that exhibit a substantially different
behavior but for which DFT-based calculations are computationally
demanding due to the large number of atoms in the unit cell. To circumvent such a limitation, the same phenomenology will be explored
within a real-space tight-binding model, that allows to ultimately consider the effects of interactions in mono-hydrogenated magic angle
superlattices with a more affordable computational cost and similar accuracy.

\section{Defect-induced states within a tight-binding model}
\label{sec:lowenergy}
We now move on to consider
the effect of a single-H atom attached to the tBLG surface
with
an effective tight-binding model 
of the form
\begin{equation}
    H_0 = 
    \sum_{\langle ij \rangle}
    t c^\dagger_i c_j
    +
    \sum_{ij}
     t_\perp (\bold r_i,\bold r_j) c^\dagger_i c_j
\end{equation}
where $\langle ij \rangle$ denotes sum over first neighbors in the same layer.
The interlayer hopping $t_\perp (\vec r)$ is modulated according to the
structure of tBLG, which is taken as
$t_{\perp}(\bold r_i,\bold r_j) = 
t_{\perp}
\frac{(z_i - z_j)^2 }{|\bold r_i - \bold r_j|^2}
e^{-\beta (|\bold r_i - \bold r_j|-d)},
$
with $\bold r_i$ the position of site $i$
and $d$ the distance between layers.
As a reference, the hopping parameters 
in graphene are $t \approx 3$ eV
and $t_\perp \approx 0.4$ eV.\cite{PhysRevB.92.075402}
For computational convenience, in the
tight-binding calculations 
an
enhanced interlayer hopping
is used,
which allows us to recover the physics
of small angle graphene superlattice
with smaller units cells.\cite{PhysRevLett.114.036601,Kaxiras18}

\begin{figure}[!t]
 \centering
	 \includegraphics[width=0.48\textwidth]{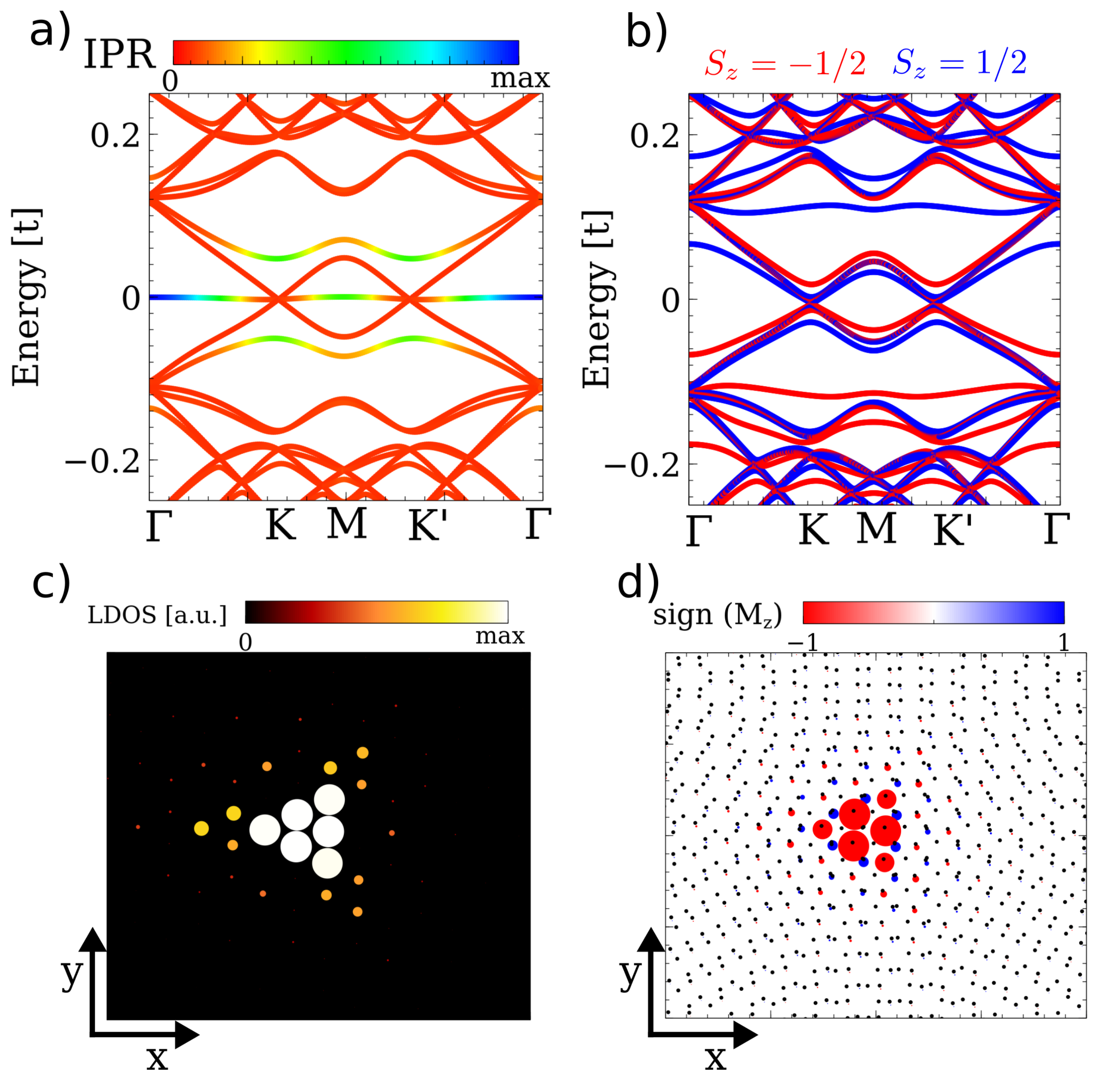}
 \caption{a) Band structure of mono-hydrogenated tBLG with a twist angle $\alpha \approx 3^\circ$, showing the emergence of a flat band with a triple-point crossing.
 In b), the self-consistent band structure exhibits the emergence of an
	exchange splitting as a consequence of interactions. 
	The
	real-space representation of the
	local density of states at Fermi level
	of panel a) is shown in c).
	The magnetism arising
	from the impurity state
	obtained after including interactions
	is shown in d).
	} 
 \label{figlargeangletb}
\end{figure}

\begin{figure}[!t]
 \centering
	 \includegraphics[width=0.48\textwidth]{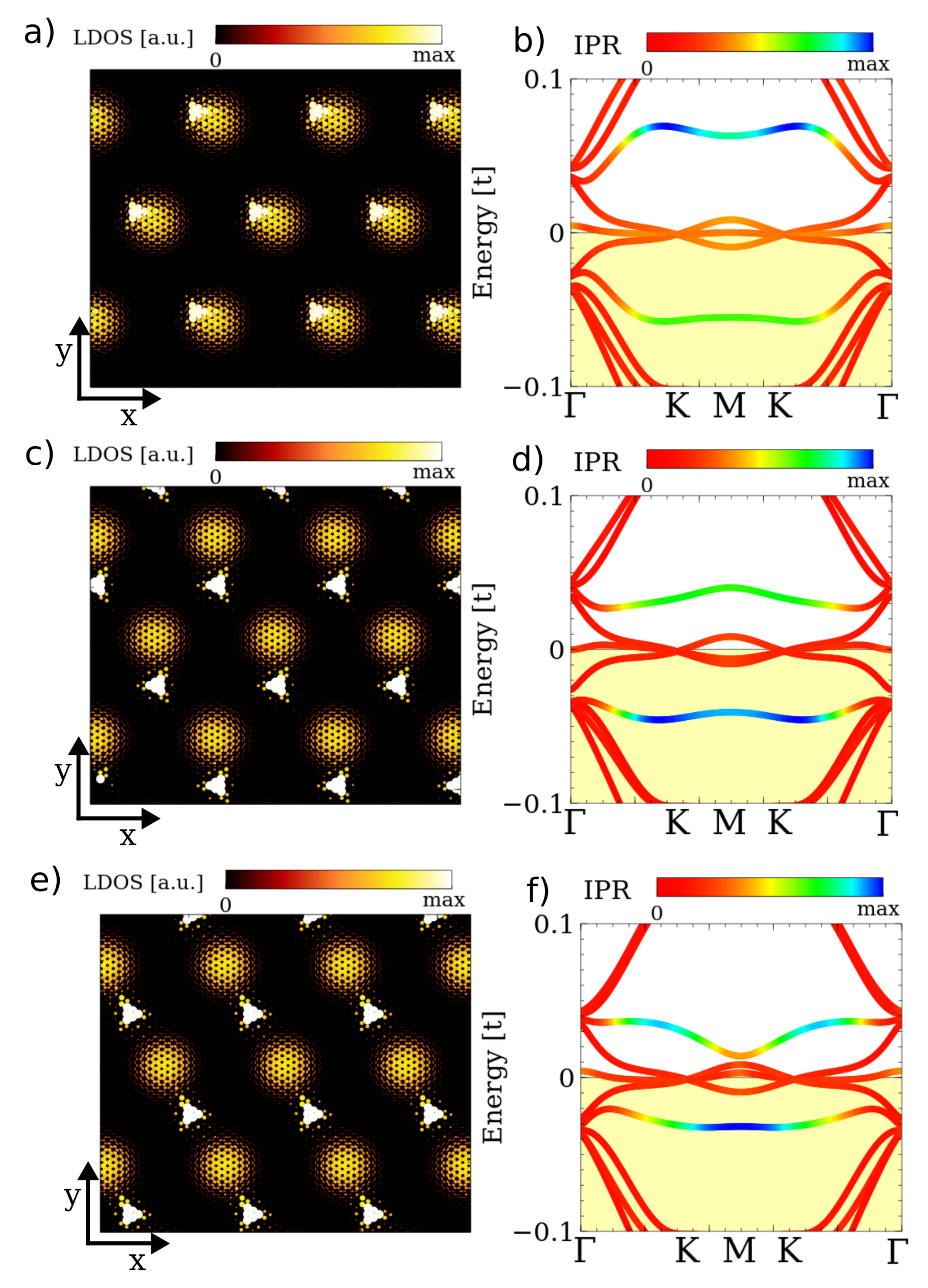}
 \caption{Local density of states (a,c,e) and band structure (b,d,f) for tBLG
	with an H atom on the surface calculated with the low-energy tight
	binding model for $\alpha \approx 1.3^\circ$.
 In a) and b) the H atom is at the AA stacking region, in c) and d) in the AB
	region, and in e) and f) in the AB/BA interface. The color code of the
	band structures (b,d,f) reflects the localization of the state in the
	supercell. A triple-point crossing at the Fermi level is observed in
	all three cases.} 
 \label{figtb}
\end{figure}

This low-energy tight-binding model aims to simplify the effect of removing a
single $p_z$ orbital by just eliminating a site out of the tBLG. As
demonstrated in the DFT calculations above, this is equivalent to attaching
a H atom to one of the bilayer surfaces.
Furthermore, in the following we will be interested in
the localization properties of the states in the moir{\'e} supercell.
This can be easily computed within the tight-binding
framework by means of
the inverse participation ratio ($IPR$) of each Bloch wave function $\Psi_n$,
that provides an estimation of the degree of
localization of a state in a moir{\'e} unit cell:
$
    IPR = \sum_i |\Psi_n(i)|^4
$.
For a fully extended state in the unit cell
the $IPR$ is a small value of $1/N$, where N is the number of atoms,
whereas for a localized state it yields
a large finite value even for large unit cells.


The emergence of magnetism is explored introducing a local Hubbard interaction term of the form
\begin{equation}
    H_U = \sum_i U 
    c^\dagger_{i,\uparrow} c_{i,\uparrow}
    c^\dagger_{i,\downarrow} c_{i,\downarrow}
    = \sum_i U n_{i,\uparrow} n_{i,\downarrow}
\end{equation}
where $n_{i,\uparrow} = c^\dagger_{i,\uparrow} c_{i,\uparrow}$
and 
$n_{i,\downarrow} = c^\dagger_{i,\downarrow} c_{i,\downarrow}$
The previous interaction is solved at the mean-field level
$H_U \rightarrow H_U^{MF} $
with 
\begin{equation}
H_U^{MF}= \sum_i U[
\langle  n_{i,\uparrow} \rangle n_{i,\downarrow} +
\langle n_{i,\downarrow} \rangle  n_{i,\uparrow} -
\langle n_{i,\uparrow} \rangle  \langle n_{i,\downarrow}\rangle
]
\end{equation}

We first address the case of a tBLG with a rotation
angle of $\alpha\approx 3^\circ$,
taking a single H atom per moir{\'e} unit cell
(Fig. \ref{figlargeangletb}).
Fig.\ref{figlargeangletb}a shows that
the paramagnetic band structure
displays a flat band meeting the linearly dispersive bands,
yielding a triple-point at
the $K$ and $K'$ points.
According to the color-resolved diagrams of the $IPR$
(Fig. \ref{figlargeangletb}a), the nearly flat band at
$E=0$ is the most localized state in the supercell, which
corresponds to the zero mode that build around the impurity
(Fig. \ref{figlargeangletb}c).
Including electronic interactions in the
model leads to a spin-dependent band
splitting
(Fig.\ref{figlargeangletb}b), 
with the consequent emergence of
magnetism
(Fig.\ref{figlargeangletb}d). 
The real-space representation of the states associated to the flat band in
Fig.\ref{figlargeangletb}c demonstrates that they are the same that become
magnetic (Fig. \ref{figlargeangletb}d).
This results exemplify
that the phenomenology of mono-hydrogenated tBLG
obtained from the DFT calculations is captured by the low-energy tight-binding
model.

We now move
on to study the fate of the impurity
bands as the magic angle regime
is approached. In
particular, at the magic angle regime,
an additional set of bands
is expected to appear, namely the
flat AA bands\cite{PhysRevB.82.121407}.
In this situation, it is expected that the impurity band will
heavily hybridize with the AA bands. 
To dive into this regime,
a single H atom in a tBLG
with twisting angle of $\alpha \approx 1.3^\circ$
(Fig. \ref{figtb}) is first considered.
As shown in Fig. \ref{figtb}a), c), and e), a set of
additional bands with highly reduced
bandwidth appear close to the charge neutrality point. 
Again, a triple point
is observed
at the $K$ and $K'$ points in the band structure,
for an impurity deposited on the AA region
(Fig. \ref{figtb}ab), 
on the AB region (Fig. \ref{figtb}cd) and in the
AB/BA boundary (Fig. \ref{figtb}ef).
Interestingly,
two additional bands appear close to charge
neutrality point, having a substantially strong $IPR$. Those
new bands appear due to the interplay
of AA band with the impurity bands, that drifts spectral
weight from the impurity state slightly above and below the
charge neutrality point.
Importantly, the small bandwidth of the AA bands suggests that 
the system may accommodate extra electrons in the AA modes,
instead of filling the impurity modes. This last feature
becomes important in the next section when dealing with
the interplay between magnetism and magic angle superconductivity.

\section{Defect-induced Yu-Shiba-Rusinov states in twisted bilayer graphene}
\label{sec:shiba}

\begin{figure}[!t]
 \centering
	 \includegraphics[width=0.48\textwidth]{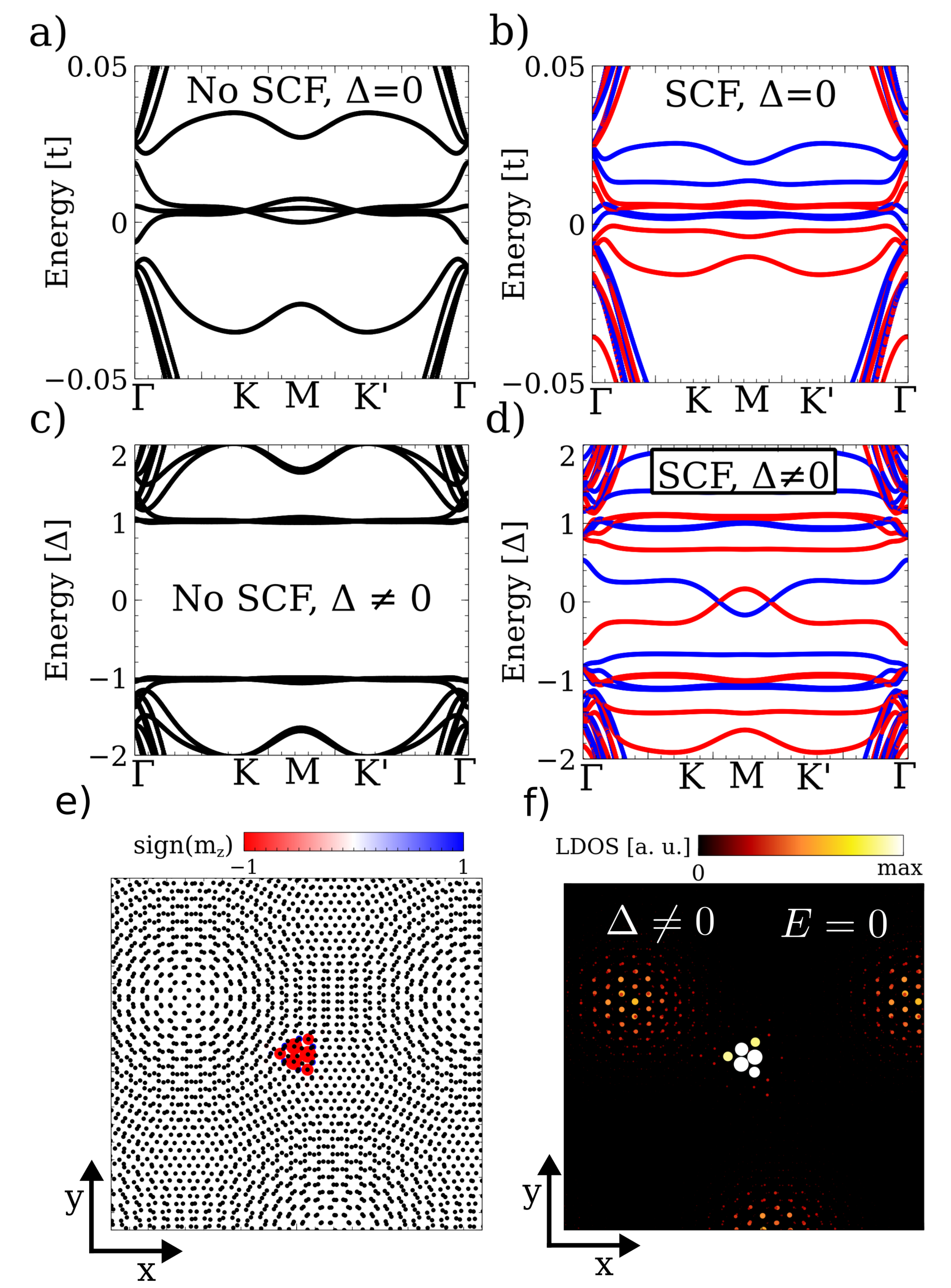}
 \caption{a) Non-interacting band structure of a single vacancy in the AB
	region for a twist angle of $\alpha\approx 1^\circ$ and a doping of
	two holes with respect to charge neutrality. b) Exchange splitting
	arises when electronic interactions are included. In c) the
	non-interacting band structure opens a gap when a pairing $\Delta$ is
	included. d) shows the combined effect of the exchange interactions and
	superconductivity is the emergence of a band with energy smaller than
	$\Delta$, signaling the emergence of Shiba-Rusinov states in the tBLG.
 Panel (e) shows the selfconsistent magnetization of (b) and (f) shows the
	local density of states of the in-gap bands of (d), highlighting that
	the in-gap Yu-Shiba-Rusinov state inherits the spatial distribution of
	the impurity induced magnetic moment.
	We took
	$U=2t$ and $\Delta=0.02t$.
	} 
 \label{figSCFSC}
\end{figure}

As the twist between the two layers
approaches a rotation angle of $\alpha\approx 1^\circ$,
the density of states of pristine twisted bilayer graphene is largely enhanced,
triggering an electronic instability
when the system is doped with nearly 2 holes per unit
cell.\cite{Cao2018,2019arXiv190306513L,Yankowitzeaav1910}
In the following, we will explore the interplay between the vacancy
induced magnetism
presented above, and the superconducting state found in magic
angle superlattices.\cite{Cao2018,2019arXiv190306513L,Yankowitzeaav1910}
To demonstrate that a magnetic defect has an impact on the tBLG superconducting
properties\cite{Cao2018,Yankowitzeaav1910,2019arXiv190306513L}, the low-energy
spectrum of a magnetic symmetry-broken tBLG is now studied
by taking the Hamiltonian
\begin{equation}
	H= H_{0} + H_{U}^{MF} + H_{SC}
\end{equation}
where $H_{SC} = \sum_i \Delta_i [c_{i,\uparrow} c_{i,\downarrow} +
c^\dagger_{i,\downarrow} c^\dagger_{i,\uparrow} ]$. 
For the sake of concreteness, we take $\Delta_i = \Delta$,
realizing a spatially uniform
s-wave superconducting singlet
pairing.\cite{PhysRevLett.121.257001,PhysRevB.98.241412,2019arXiv190200763P,2018arXiv180704382L}
We point out that additional superconducting symmetries
have been proposed for magic
angle superlattices besides the one
consider above.\cite{PhysRevLett.121.087001,PhysRevX.8.041041}

We first consider the case without superconducting order,
namely $\Delta=0$.
The non-interacting band structure for an angle $\alpha \approx 1^\circ$ tBLG
mono-hydrogenated in the AB region is shown in Fig.\ref{figSCFSC}a. 
The triple-point survives in this small angle regime and the flat band is
strongly hybridized with the nearly flat Dirac cones, located in the AA region. 
In addition, new
bands appear close to charge
neutrality carry spectral weight of the impurity state. 
The fate of the impurity-induced
magnetic state triggered by electron interactions is considered next.
When introducing interactions and
doping with two holes
per moir{\'e} unit cell,
the impurity induced
magnetic moment survives (Fig.\ref{figSCFSC}b) and e)).
This strikingly compares with the large angle
scenario, in which doping with 
a single electron per moir{\'e} unit cell already
destroys the impurity-induced magnetism.

We now unveil the effect of superconductivity in the previous setup.
For that sake, a
superconducting term is included on top of the
normal state Hamiltonian.
In a system with time-reversal symmetry and uniform pairing, the BdG
eigenvalues $\epsilon_k$ are given by $\epsilon_k = \pm \sqrt{E^2_k +
\Delta^2}$, where $E_k$ are the Bloch eigenvalues in the absence of
superconductivity. Namely, the BdG band structure of a nonmagnetic system shows
a gap of $2\Delta$ in the spectra.
This is observed in Fig.\ref{figSCFSC}c,
where the superconducting field was included on top of the
paramagnetic non-interacting band
structure. 

In the presence of magnetic moments, states
inside the superconducting gap can appear
due to the breaking of time reversal symmetry.
In particular,
Yu-Shiba-Rusinov states\cite{RevModPhys.78.373,PhysRevLett.117.186801,PhysRevLett.120.156803,PhysRevLett.115.087001,PhysRevB.78.035414} are
in-gap modes in the superconducting spectrum that appear as a result of the
time-reversal symmetry breaking introduced by local magnetic moments. 
Adding a uniform superconducting pairing $\Delta$ to the
self-consistent solution of Fig.\ref{figSCFSC}b, two in-gap bands are observed
in the BdG spectrum, as shown in Fig.\ref{figSCFSC}d. 
Interestingly, the Yu-Shiba-Rusinov state
(Fig. \ref{figSCFSC}f)
reflects the spatial structure of the magnetic
state
(Fig. \ref{figSCFSC}e).
\cite{PhysRevLett.117.186801,PhysRevLett.120.156803,PhysRevLett.115.087001}
This results highlight that the impurity-induced
magnetic moment of the tBLG can coexist with the superconducting state,
generating in-gap Yu-Shiba-Rusinov states that could
be observed by means of scanning tunnel microscope.\cite{PhysRevLett.117.186801,PhysRevLett.120.156803,PhysRevLett.115.087001}

Finally, it is interesting to note that the energy of this Yu-Shiba-Rusinov
states cannot be easily estimated as
in conventional metals hosting magnetic impurities.\cite{RevModPhys.78.373}
In those instances,
the energy of the in-gap state generated in the superconducting states takes the form\cite{Shiba1968,RevModPhys.78.373}
$
\epsilon = 
\Delta
\frac
{1-(JS\pi\rho/2)^2}
{1+(JS\pi\rho/2)^2}
$
where $J$ is the exchange coupling between the impurity and the conduction
electrons, $S$ is the spin of the impurity, and $\rho$ is the density of states
in the normal state.
In mono-hydrogenated tBLG the density of states in the normal state is
divergent due to the existence of nearly flat bands, and the onset of magnetism
changes substantially the density of states and the Fermi energy, invalidating
the previous procedure. This phenomenology is also found in hydrogenated.
monolayer graphene\cite{Lado2016} Nevertheless, the emergence of
Yu-Shiba-Rusinov states is a robust feature, as shown in the exact solution
presented above.

\section{\label{sec:conclusions} Conclusions}
To summarize, we have presented
a detailed description of
the impact of two types defects
on the electronic and magnetic properties 
of twisted bilayer graphene.
Localized states are found for both a vacancy and
mono-hydrogenation of tBLG, which yield spatially localized magnetic states.
Both types of defects remove effectively a p$_z$ electron out of the bilayer.
On one hand, mono-hydrogenation defect states modify the low-energy band
diagram at the charge neutrality point creating a triple-point.
On the other hand,
a carbon vacancy yields impurity bands
at different energies, distorting the $\pi$ bands at the Fermi level.
Interestingly, the presence of impurities leads to the
emergence of spatially localized
magnetic moments whose magnitude depends on the
location of the defect.

In the magic angle regime, doping with two holes per moir{\'e} unit cell
does not destroy the hydrogen induced magnetic moment, suggesting that such  a
defect-induced magnetism may coexist with the superconducting state.
Including a superconducting term in the Hamiltonian, the interplay between
defect-induced magnetism and superconductivity gives rise to in-gap
Yu-Shiba-Rusinov states in the twisted bilayer.
Our results demonstrate that atomic imperfections may have a substantial impact in
the spectral properties of superconducting twisted graphene superlattices.

\section*{\label{acknowledgments}Acknowledgments}
Los Alamos National Laboratory is managed by Triad National Security, LLC, for
the National Nuclear Security Administration of the U.S. Department of Energy
under Contract No. 89233218CNA000001. This work was supported by the U.S. DOE
Office of Basic Energy Sciences Program E3B5.
We thank I. Brihuega, E. Cortes-del Rio and A. Ramires for
fruitful discussions.
A.L.-B. acknowledges the
computing resources provided on Bebop, the high-performance computing clusters
operated by the Laboratory Computing Resource Center at Argonne National
Laboratory. J. L. L. acknowledges financial support from the ETH Fellowship
program. 

\appendix

\section*{Appendix}

\section{\label{sec:methods } First-principles calculations}
The description of the coupling between the graphitic layers before and after
the introduction of an impurity was conducted through self-consistent
calculations with the SIESTA code within a localized orbital basis set scheme.
Spin-polarized calculations were conducted using a double-$\zeta$ basis set,
and the local density approximation (LDA) approach for the exchange-correlation
functional was used. Atomic positions of systems formed by over 508 atoms were
fully relaxed with a force tolerance of 0.01 eV/\AA. The integration over the
Brillouin zone (BZ) was performed using a Monkhorst sampling of 7 $\times$ 7
$\times$ 1 k-points. The radial extension of the orbitals had a finite range
with a kinetic energy cutoff of 50 meV. A vertical separation of 25 \AA\ in the
simulation box prevents virtual periodic parallel layers from interacting.

\end{document}